\documentclass[10pt,twocolumn]{article}
\usepackage{graphicx}
\usepackage{times}

\title{A Game Theoretic Framework for Incentives in P2P Systems
\thanks{The authors were  supported in part by National
Science Foundation grants IIS-0121562 (C.B and S.S) and 9970700 (D.A)}}
\author{Chiranjeeb Buragohain \and
  Divyakant Agrawal \and Subhash
  Suri\\
Computer Science Department\\University of California,
Santa Barbara, CA 93106 \\
\texttt{\{chiran,agrawal,suri\}@cs.ucsb.edu}}
\date{}
\begin{document}
\maketitle

\begin{abstract}
Peer-To-Peer (P2P) networks are self-organizing, distributed systems, 
with no centralized authority or infrastructure. Because of the voluntary
participation, the availability of resources in a P2P system can be 
highly variable and unpredictable. In this paper, we use ideas from
\emph{Game Theory} to study the interaction of strategic and rational 
peers, and propose a \emph{differential service-based incentive scheme}
to improve the system's performance.
\end{abstract}

\section{Introduction}

Peer-To-Peer (P2P) systems are self-organizing, distributed
resource-sharing networks. They differ from traditional distributed 
computing systems in that no central authority controls or manages the 
various components; instead, nodes form a dynamically changing and 
self-organizing network. By pooling together the resources of many 
autonomous machines, P2P systems are able to provide an inexpensive 
platform for distributed computing, storage, or data-sharing that 
is highly scalable, available, fault tolerant and robust.
As a result, a large number of academic and commercial projects are
underway to develop P2P systems for various applications
\cite{gnutella, kazaa, pastry, can,chord,freenet}.

The democratic (or anarchic) nature of P2P systems, which is
responsible for their popularity and scalability, also has serious potential
drawbacks. There is no central authority to mandate or coordinate the 
resources that each peer should contribute. Because of the voluntary 
participation, the system's resources can be highly variable and 
unpredictable.  Indeed,
%
in a recent experimental study of Napster and Gnutella, 
Saroiu et~al.~\cite{saroiu} found that many users are simply consumers, and 
do not contribute much to the system.
In particular, they found that
(1) user sessions are relatively short;  $50\%$ of the sessions are shorter 
than 1 hour, and 
(2) many users are \emph{free riders}; that is, they contribute little or 
nothing. For example, in the Gnutella system, 25\% of the users share no files
at all.

Short sessions mean that a significant portion of the data in the system 
might be unavailable for large periods of time---the hosts with those
data are offline. Short uptimes also hurt system performance because there 
are fewer servers to download files from.
Similarly, as a growing number of users become free riders, the system 
starts to lose its peer-to-peer spirit, and begins to resemble a more 
traditional client-server system.

If the P2P systems are to become a reliable platform for distributed
resource-sharing (storage, computing, data etc), then they must
provide a predictable level of service, both in content and performance.
A necessary step towards that goal is to develop mechanisms by
which contributions of individual peers can be solicited and predicted.
In a system of autonomous but \emph{rational} participants, a reasonable
assumption is that the peers can be incentivized using economic principles.
Two forms of incentives have been considered in the past \cite{golle}:
(1) monetary payments (one pays to consume resources and is paid to
	contribute resources), and
(2) differential service (peers that contribute more get better
	quality of service).
The monetary payment scheme involves a fictitious currency, and
requires an accounting infrastructure to track various resource
transactions, and charges for them using micropayments.  While the
monetary scheme provides a clean economic model, it seems highly
impractical. For instance, see~\cite{odlyzko} for arguments against
such a scheme for network pricing.

The differential service seems more promising as an incentive model,
and that is the direction we follow. There are many different ways to
differentiate among the users.  For instance, one could define a
\emph{reputation index} for the peers, where the reputation reflects a
user's overall contribution to the system. In fact, a reputation based
mechanism is already used by the KaZaA \cite{kazaa} file sharing
system; in that system, it's called the \emph{participation level}.
Quantifying a user's reputation and prevention of faked reputations,
however, are thorny problems.

In general, since the nodes in a P2P systems are strategic players,
they are likely to manipulate any incentive system.
As a result, we argue that a correct tool for modeling the interaction of 
peers is \emph{game theory} \cite{fudenberg}.
We introduce a formal model of \emph{incentives through differential 
service} in P2P systems, and use the game theoretic notion of 
\emph{Nash Equilibrium} to analyze the strategic choices by 
different peers.



We treat each peer in the system as a rational, strategic player,
who wants to maximize his utility by participating in the P2P system.
The utility of a peer depends on his benefit (the resources of 
the system he can use) and his cost (his contribution). 
Our \emph{differential service} model links the benefit any peer
can draw from the system to his contribution---the benefit is
a monotonically increasing function of a peer's contribution.
Thus, this is a \emph{non-cooperative} game among the peers:
each wants to maximize his utility. The classical concept of Nash
Equilibrium points a way out of the endless cycle of speculation and 
counter-speculation as to what strategies the other peers will use.
An equilibrium point is a \emph{locally optimum} set of strategies 
(contribution levels in our case), where no peer can improve
his utility by deviating from the strategy.
While Nash equilibrium is a powerful concept, computing these
equilibria is not trivial. In fact, no polynomial time algorithm is
known for finding the Nash equilibrium of a general $N$ person game.

We first consider a simplified setting, \emph{homogeneous peers},
where we assume that all peers derive equal benefit from everybody else 
(homogeneity of peers).
In this case, we show
(1) there are exactly two Nash equilibria, and
(2) there are closed-form analytic formulae for these equilibria.
We also investigate the stability properties of these equilibria,
and show that in a repeated game setting, the equilibrium with 
the better system welfare will be realized.

We next consider the case of \emph{heterogeneous peers}, 
where the interaction matrix is an arbitrary $N \times N$ matrix.
That is, we allow an arbitrary benefit function for each pair of peers. 
No closed form solution is possible for this setting, and
so we study this using simulation.
We use the homogeneous case as a benchmark to see how well
the simulation tracks the theoretical prediction.
Our main findings are that the \emph{qualitative} properties of the 
Nash equilibrium are impervious to 
(1) exact form of the probability function used to implement differential
	service,
(2) perturbations like users leaving and joining the system,
(3) non-strategic or non-rational players, who do not play according to
	the rules, etc. 
Finally, we discuss practical ways of implementing a differential service 
incentive scheme in a P2P system.

\section{Our Incentive Model}

\subsection{Strategy and Nash Equilibrium}
\label{game-sec}

A traditional distributed system assumes that all participants in the
system work together cooperatively; the participants in the system
share a common goal, do not compete with each other or try to
subvert the system.  A P2P systems, on the other hand, consists of 
autonomous components: users compete for shared but limited
resources (e.g. download bandwidth from popular servers) and, at 
the same time, they can restrict the download from their own 
server by denying access or not contributing any resources.  
As such, the interaction of the various peers in a P2P system is best modeled 
as a \emph{non-cooperative game} among rational and strategic players.
The players are rational because they wish to maximize their own 
gain, and they are strategic because they can choose their actions
(e.g. resources contributed) that influence the system.
%
The behavior that a player adopts while interacting with other players 
is known as that player's \textit{strategy}.  In our setting, a peer's 
strategy is his level of contribution.
%
%
The player derives a benefit  from his interaction with 
other players which is termed as a payoff or \textit{utility}.  
Interesting economic
behavior occurs when the utility of a player depends not only on his
own strategy, but on everybody else's strategy as well.  The most
popular way of characterizing this dynamics is in terms of
\textit{Nash equilibrium}.  Since the utility or payoff of a player is
dependent on his strategy, he might decide to unilaterally switch his
strategy to improve his utility.  This switch in strategy will affect
other players by changing their utility and they might decide to switch
their strategy as well.  The collection of players is said to be at
Nash equilibrium if no player can improve his utility by unilaterally
switching his strategy.  
In general, a system can have multiple equilibria.


\subsection{Incentives and Strategies in P2P System}
\label{incentive-sec}

We assume that there are $N$ users (peers) in the system,
$P_1, P_2, \ldots, P_N$.
We will denote the utility function of the $i$th peer as $U_i$.
This utility depends on several parameters which we shall discuss below 
one by one.

\subsubsection{Measuring the Contribution}  

We will use a single number $D_i$ to denote the contribution of $P_i$.
The precise definition of $D_i$ is immaterial as long as it can
be quantified and treated as a continuous variable.  
For concreteness, we will take $D_i$ to be the \emph{cumulative disk space}:
disk space contribution integrated over a fixed period of time, say a week.
One can also use other metrics such as number of downloads served by this
peer to other peers.

For each unit of resource contributed, the peer incurs a cost $c_i$ 
(measured in dollars).  So the total cost of $P_i$ for participating in the
system is $c_iD_i$.  We shall  find it convenient to define a
dimensionless contribution

\begin{equation}
  d_i ~\equiv~ D_i/D_0,
\end{equation}
where $D_0$ is an absolute measure of contribution (say 20MB/week).
$D_0$ is a constant that the system architect is free to set---our incentive 
scheme will strive to ensure that all peers make a contribution at least $D_0$.

\subsubsection{The Benefit Matrix} 

Each peer's contribution to the system potentially benefits all other
peers, but perhaps to varying degrees. We encode this benefit using 
a $N \times N$ matrix $B$, where $B_{ij}$ denotes how much the contribution 
made by $P_j$ is worth to $P_i$ (measured in dollars). For instance, if 
$P_i$ is not interested in $P_j$'s contribution, then $B_{ij} = 0$. 
In general, $B_{ij} \geq 0$, and we assume that $B_{ii} = 0$, for all $i$.
Again, we define a set of dimensionless parameters corresponding to
$B_{ij}$ by

\begin{equation}
  b_{ij} = B_{ij}/c_i, \;\; b_i = \sum_j b_{ij},\;\; b_\mathrm{av} =
  \frac{1}{N}\sum_i b_i
\end{equation}

$b_i$ is the total benefit that $P_i$ can derive from the system if
all other users make unit contribution each.  $b_i$ will turn out to
be an important parameter in determining whether it is worthwhile for
$P_i$ to join the system.  We shall show that there exists a critical
value of benefit $b_c$ such that if $b_i < b_c$, then $P_i$ is better
off not joining the system.  $b_\mathrm{av}$ is simply the average of
$b_i$ for the whole system.

\subsubsection{Probability as Service Differentiator}

The differential service is a game of expectations: a peer rewards
other peers in proportion to \emph{their contribution}.
A simple scheme to implement this idea is as follows:
peer $P_j$ accepts a request for a file from peer $P_i$ with
probability $p(d_i)$, and rejects it with probability $1 - p(d_i)$.
Thus, if $P_i$'s contribution is small, its request is more
likely to be rejected.
There are many enhancements and improvements to this simple idea.
One could, for example, curtail the search capabilities of a peer 
depending on his contribution. In the Napster model, one could return 
only a fraction $p(d_i)$ of the total results found.  
We also assume that every request from peer $P_i$ is tagged with his 
contribution $d_i$ as metadata.
We will discuss some of these enhancements and implementation
issues in section \ref{impl-sec}.

\begin{figure}
  \begin{center}
    \includegraphics[width=0.35\textwidth]{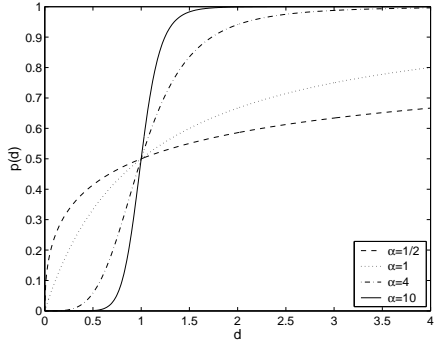}
    \caption{$p(d)$ plotted as a function of $d$ for values of
    $\alpha$ 1/2, 1, 4 and 10.}
    \label{prob-fig}
  \end{center} 
\end{figure}

It turns out that the choice of the exact probability function 
does not affect the \emph{qualitative} nature of our results.
Any reasonable probability function that is a monotonically
increasing function of the contribution should do.
In our analysis, we have chosen the following natural form:

\begin{equation}
  p(d) = \frac{d^\alpha}{1+d^\alpha},\;\; \alpha > 0.
\label{probfn}
\end{equation}
It has the desirable properties that $p(0) = 0$, and $p(d) \rightarrow 1$
as $d$ gets large.
The choice of the exponent $\alpha$ determines how ``step-function-like''
the probability function is. 
See Figure~\ref{prob-fig}.
For small values, say $\alpha = 1$, the function is rather smooth; 
but for larger values, say $\alpha = 10$, the function has a steep step;
for contribution below the step, requests have high probability of
rejection; and 
for contribution above the step, requests have high probability of
acceptance.

\subsubsection{The Utility Function} 

With these cost and benefit parameters, the total utility $U_i$ that
$P_i$ will derive by joining the system is
\begin{equation}
  U_i = -c_iD_i + p(d_i)\sum_jB_{ij}D_j,\; B_{ii} \equiv 0
\end{equation}
The first term is the cost to join the system, while the second
term is the total expected benefit from joining the system.  In terms
of the dimensionless parameter
\begin{equation}
u_i = \frac{U_i}{c_iD_0}
\end{equation}
we rewrite the utility as
\begin{equation}
  u_i = -d_i + p(d_i)\sum_jb_{ij}d_j, \; b_{ii} \equiv 0
\label{model}
\end{equation}
The $-d_i$ term is simply $P_i$'s cost to join the system and it increases
linearly as $P_i$ contributes more disk/bandwidth to the system.  
$P_i$'s benefit depends on how much the other peers are contributing to the 
system ($d_j$), what that contribution is
worth to him ($b_{ij}$), and how probable it is that he will be able to
download that content ($p(d_i)$).  Using the fact that $p(0) = 0$ and
$p(\infty) = 1$, we can find the two limits of the utility function :
  \begin{equation}
    \lim_{d_i\rightarrow 0} u_i  =  0, \;\;\; 
    \lim_{d_i\rightarrow\infty} u_i  =  -\infty.
  \end{equation}
  \begin{figure}
    \begin{center}
      \includegraphics[width=0.35\textwidth]{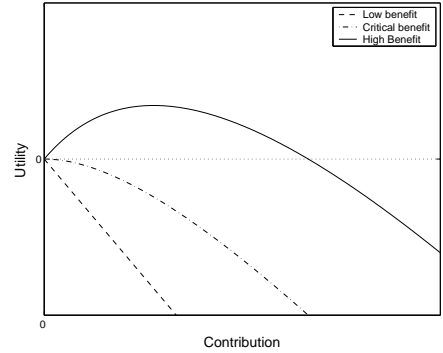}
      \caption{A qualitative plot of utility vs. the contribution/strategy;
	the scales on X and Y axes are arbitrary.  Unless there is a 
	critical level of benefit $b_c$,
	the utility for the peer is always less than 0.}
      \label{utility-fig}
    \end{center}
  \end{figure}

Thus, neither extreme maximizes a peer's utility.
The value of an intermediate strategy depends on the
contribution of other users and the worth of those contributions.
See Figure~\ref{utility-fig} for a graphical representation of a possible
utility function for different levels of benefit $b_i$.  If $b_i$
exceeds a critical value $b_c$, then it is possible for the utility
function to have a maximum and only then the peer would want to join
the system.

In the next section we start with the discussion of Nash equilibrium
for the model that we have just described.

\section{Nash Equilibrium in the Homogeneous System of Peers}

We define a homogeneous system of peers to be a system where $b_{ij} =
b$ for all $i\neq j$; in other words in this system all peers derive
equal benefit from everybody else.  This simplified system allows us to 
study the problem in an idealized setting, and gain insights that
can be applied to the more complex heterogeneous system. In the
homogeneous system, the model of equation \ref{model} reduces to

\begin{equation}
  u = -d + p(d)(N-1)bd.
\label{hom-eqn}
\end{equation}

$b_i = b_\mathrm{av} = b(N-1)$ for all peers $P_i$.
By symmetry, therefore, the problem reduces to a 2-person game,
which we analyze below.

\subsection{The Two Player Game}

In a homogeneous system of two players, Equation \ref{model} reduces
to
\begin{eqnarray}
  u_1 & = & -d_1+b_{12}d_2p(d_1) \nonumber \\ 
  u_2 & = & -d_2+b_{21}d_1p(d_2)
\label{two-eqn}
\end{eqnarray}
For algebraic simplicity, let us also assume that $\alpha=1$,
i.e. $p(d) = d/(1+d)$.  As discussed in section \ref{incentive-sec},
we expect that if the benefits that the peers derive from each other,
i.e. $b_{12}$ and $b_{21}$ are too small then it will be best for the
peers not to join.  The question to ask at this point is whether a
Nash equilibrium exists for large enough values of benefits where both
peers can derive non-zero utility from their interaction.   

This model is very similar to the Cournot duopoly model
\cite{fudenberg} and we can analyze it using similar methodology.
Suppose $P_2$ decides to make a contribution $d_2$ to the system.
Given this contribution $d_2$, naturally the best thing for $P_1$ to
do is to tune his $d_1$ such that it maximize his utility $u_1$.
Maximizing $u_1$ with respect to $d_1$, we immediately find that the
best response $d_1$ is given by
\begin{equation}
r_1(d_2) \equiv   d_1 = \sqrt{b_{12}d_2} - 1,
\label{rxn1}
\end{equation}
where $r_1(d_2)$ is known as the \textit{reaction function} for $P_1$.
This is the best reaction for $P_1$, given a fixed strategy for $P_2$.
Since $P_2$ knows that $P_1$ is going to respond in this fashion, his
own reaction function to 1's strategy is
\begin{equation}
r_2(d_1) \equiv   d_2 = \sqrt{b_{21}d_1} -1.
\label{rxn2}
\end{equation}
Nash equilibrium 
\footnote{For readers versed in game theory, we want to say that we
 are only interested in pure strategy Nash equilibrium.  A mixed
 strategy will correspond to a peer probabilistically choosing a
 contribution.  Such a scenario is inadmissible and and we shall not
 discuss it any further} 
exists if there is a set of $(d_1^*,
 d_2^*)$, such that they form a fixed point for equations \ref{rxn1}
 and \ref{rxn2}, i.e. the fixed points satisfy
\begin{eqnarray}
  d_1^* & = & \sqrt{b_{12}d_2^*} -1, \nonumber\\
  d_2^* & = & \sqrt{b_{21}d_1^*} -1.
\label{fxdpt}
\end{eqnarray}
Finding the fixed point is much easier if we assume $b_{12} = b_{21} =
b$ (this is the homogeneous peer system).  In that case $d_1^* = d_2^*
= d^*$ and the solution of equation \ref{fxdpt} is
\begin{equation}
  d^* = (b/2-1) \pm \left((b/2-1)^2-1\right)^{1/2}
\label{eqbm-contrib-eqn}
\end{equation}
\emph{A solution to this equation exists only if $b\geq 4 \equiv
  b_c$.}
  \footnote{
For general values of $\alpha$,
\begin{equation}
  d^* = \left((b\alpha/2-1) \pm
  \left((b\alpha/2-1)^2-1\right)^{1/2}\right)^{1/\alpha}
\label{eqbm-contrib-alpha-eqn}
\end{equation}
A solution to this equation exists only if $b\alpha \geq 4$.}
Thus,  $b_c = 4$ is the critical value of benefit illustrated in Figure~
\ref{utility-fig} below which it is not profitable for a peer to join
the system.  Note that this critical value $4$ is an artifact of the
form of the $p(d)$ we chose.  For different choices of $p(d)$, this
constant $b_c$ will change, but will always be a
constant independent of the number of peers in the system.  For $b =
b_c$, the only solution is $d_1^* = d_2^* = 1$.  For $b > b_c$, there are
two solutions
\begin{equation}
  d_1^*=d_2^*=d^*_\mathrm{lo} < 1, \; \mathrm{and} \; d_1^*=d_2^*=d^*_\mathrm{hi} > 1.
\end{equation}
\begin{figure}
  \begin{center}
    \includegraphics[width=0.35\textwidth]{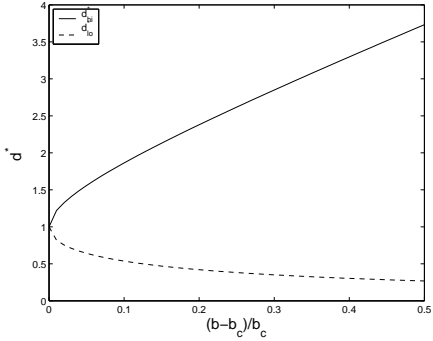}
    \caption{The Nash equilibrium contributions for the two peer
      system plotted as a function of scaled benefit $(b-b_c)/b_c$.
      For $b < b_c$, there are no equilibria.  For all $b > b_c$
      there are two possible equilibria }
    \label{nash-fig}
  \end{center}
\end{figure}
which are plotted in Figure~\ref{nash-fig}

\subsection{The $N$ Player Game}
At this point we can come back to the homogeneous system of peers of
equation \ref{hom-eqn}.  A comparison of equations \ref{hom-eqn} and
\ref{two-eqn} shows that for the homogeneous system of peers, the
fixed point equations \ref{fxdpt} are now
\begin{equation}
  d^* = \sqrt{b(N-1)d^*}-1,
\end{equation}
or in other words
\begin{equation}
  d^* = (b(N-1)/2-1) \pm \left((b(N-1)/2-1)^2-1\right)^{1/2}.
\label{sym-equilibrium}
\end{equation}
So, with the replacement of $b$ by $b(N-1)$, the results for the two
peer system are exactly applicable for the $N$ player system as well.
Although the homogeneous peer system is not realistic, we shall see
that the \textit{average} properties of the Nash equilibria for the
heterogeneous system closely follow the homogeneous case.

\subsection{Stability of the Nash Equilibria}
\label{equilibrium-prop-sec}

\begin{figure}
  \begin{center}
    \includegraphics[width=0.35\textwidth]{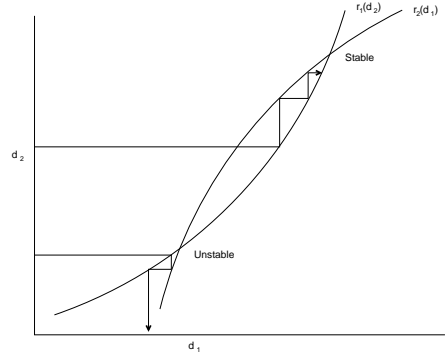}
    \caption{The Cournot learning process near the vicinity of the two
    fixed points.  Here we have plotted the reaction functions from
    equations \ref{rxn1},\ref{rxn2}.  For any starting value of $d_2 >
    d^*_\mathrm{lo}$ ($d^*_\mathrm{lo}$ is the unstable fixed point),
    the learning process converges to the stable fixed point.  If the
    starting point is too close to the origin, then the iteration
    moves away from the unstable fixed point and eventually ends up at
    0.}
    \label{learning-fig}
  \end{center}

\end{figure}

Since our system has two possible Nash equilibria, the natural
question arises which equilibrium will be chosen by the system in
practice.  There is a natural learning scenario between peers which
can help us answer this question.
Suppose  the user $P_2$ sets his contribution to some
$d_2$ to start with.  In this situation, $P_1$ can use the reaction function
$r_1(d_2)$ to set his optimum contribution at $d_1$.  Seeing this
contribution $P_2$ adjusts his own contribution and thus each peer
takes turns in setting their contribution.  If this process converges,
then naturally that level of contribution for $P_1$ and $P_2$ will
constitute a Nash equilibrium, i.e.
\begin{eqnarray}
  d_1^* & = & r_1(r_1(r_1(r_1(.....(d_2))))) \nonumber \\
  d_2^* & = & r_2(r_2(r_2(r_2(.....(d_1))))).
\label{iterations}
\end{eqnarray}
The learning process and convergence is graphically outlined in
Figure~ \ref{learning-fig}.  From the figure we see that under this
learning process, either the peers will quit the game (zero utility)
or they will converge to the equilibrium $d_\mathrm{hi}^*$.  Note that
this iterative procedure gives us an algorithm to find the stable Nash
equilibrium of a game and we shall make use of it in section \ref{hetero-sec}.
The fixed point $d^*_\mathrm{hi}(d^*_\mathrm{lo})$ is  \emph{locally
stable(unstable)}, i.e. if the two peers start near the fixed point,
under iteration of the mappings, they will move closer to (away from)
the fixed point.  It is gratifying to see that the stable Nash
equilibrium $d^*_\mathrm{hi}$ is also the desirable equilibrium for
the performance of the system.

The stability of the fixed points can be estimated by linearizing the
mappings $r_1$ and $r_2$ near the fixed point \cite{moulin}.  Consider
a point $(d^*_1+\delta d_1, d^*_2+\delta d_2)$ close to the fixed
point $(d^*_1, d^*_2)$.  Expanding equation \ref{fxdpt} around the
fixed point, we find that after one iteration, the new deviations are
given by
\begin{equation}
  \left[
    \begin{array}{c}
      \delta d_1^\prime\\
      \delta d_2^\prime
    \end{array}
    \right] = 
  \left[
    \begin{array}{cc}
      0 & (d_1^*+1)/(2d_2^*)\\
      (d_2^*+1)/(2d_1^*) & 0
    \end{array}
    \right]
  \left[
    \begin{array}{c}
      \delta d_1\\
      \delta d_2
    \end{array}
    \right].
\label{perturbation}
\end{equation}
The new deviation will be smaller in magnitude than the old deviation
provided the maximum eigenvalue
$\sqrt{(d_1^*+1)(d_2^*+1)/(4d_1^*d_2^*)}$ of the matrix on the RHS is
smaller than 1.  For $b > b_c$, the fixed point ($d^*_\mathrm{hi} >1
$), is stable and the other fixed point ($d_\mathrm{lo}^*$) is
unstable.  For $b=b_c$, the two fixed points collapse into one.  The
eigenvalues of the matrix are exactly equal to one and the deviations
neither increase, nor decrease in magnitude, i.e. the fixed point is
\emph{neutral}.



\section{Nash Equilibrium in the  Heterogeneous System of Peers} 
\label{hetero-sec}
In a heterogeneous system, we need to deal with the full complexity of
the model.  The fixed point equations for $\alpha=1$ can be
immediately derived in analogy with the two player game (equation
\ref{fxdpt}) as
\begin{equation}
  d_i^* = \left[\sum_{j\neq i} b_{ij}d_j^*\right]^{1/2} -1.
\label{het-eqn}
\end{equation}
Since it is not possible to solve this set of equations analytically,
we use an iterative learning model to solve this system of equations.

\subsection{The Learning Model and Simulation Results}
Let us consider the interaction of users in a real P2P system.  Any
particular peer $P_i$ interacts only with a limited set of all
possible peers --- these are the peers who serve files of interest to
$P_i$.  As it interacts with these peers, $P_i$ learns of  the
contributions made by them and to maximize its utility adjusts its own
contribution.  Obviously this contribution that $P_i$ makes is not
globally optimal because it is based only on information from a
limited set of peers.  But after $P_i$ has set its own contributions,
this information will be propagated to the peers it interacts with and
those peers will adjust their own contribution.  In this way the
actions of any peer $P_i$ will eventually reach all possible peers.
The reaction of the peers to $P_i$'s contribution will affect $P_i$
itself and it will find that perhaps it will be better off by
adjusting its contribution once more.  In this way, every peer will go
through an iterative process of setting its contribution.  If and when
this process converges, the resulting contributions will constitute a
Nash equilibrium.

The iterative learning algorithm that we have chosen to solve equation
\ref{het-eqn} mimics this learning process.  To start with, all the
peers have some random set of contributions.  In a single iteration of
the algorithm, every peer $P_i$ determines the optimal value of $d_i$
that it should contribute given the values of $d$ for other peers and
the values of $b_{ij}$.  At the end of the iteration the peers update
their contribution to their new optimal values.  Since now the
contributions $d_i$ are all different, the peers need to recompute
their optimal values of $d_i$ and we can start the next iteration.
When this iterative process converges to a stable point, we reach a
Nash equilibrium.  In the following numerical experiments we
demonstrate that for heterogeneous system of peers, the iterative
learning process does converge to the desirable Nash equilibrium
$d^*_\mathrm{hi}$ and we compare the results with the analytic results
for the system of homogeneous peers.

\subsubsection{Choice of Parameters}
We choose the number of peers $N$ to be from 500-1000.  Since a peer
$P_i$ interacts only with a small subset of its peers, $b_{ij}$ is
non-zero only for a few values of $j$.  We also assume that the peers
for which $b_{ij}$ is non-zero are picked randomly from all possible
peers.  Note that this subset is not the set of neighbors in the
overlay network sense, but the set of other peers with whom it
exchanges files.  The size of the set for which $b_{ij} \neq 0$ is
chosen to be 2\% of $N$.  In general for smaller value of this
fraction, the algorithm takes longer to reach the Nash equilibrium,
but the equilibrium itself does not change.  The values of $b_{ij}$ do
not evolve in time and we choose them from a Gamma distribution.  The
choice of Gamma distribution was arbitrary, we have experimented with
Gaussian distribution as well.  We choose the initial values of $d_i$
from a Gaussian distribution.  
The distribution $d_i$ evolves at every iteration
and eventually converges to the Nash equilibrium distribution.  The
value of $\alpha$ for all our results is 1.0 unless otherwise
specified.

\subsubsection{Convergence to Nash Equilibrium}
\begin{figure}
  \begin{center}
    \includegraphics[width=0.35\textwidth]{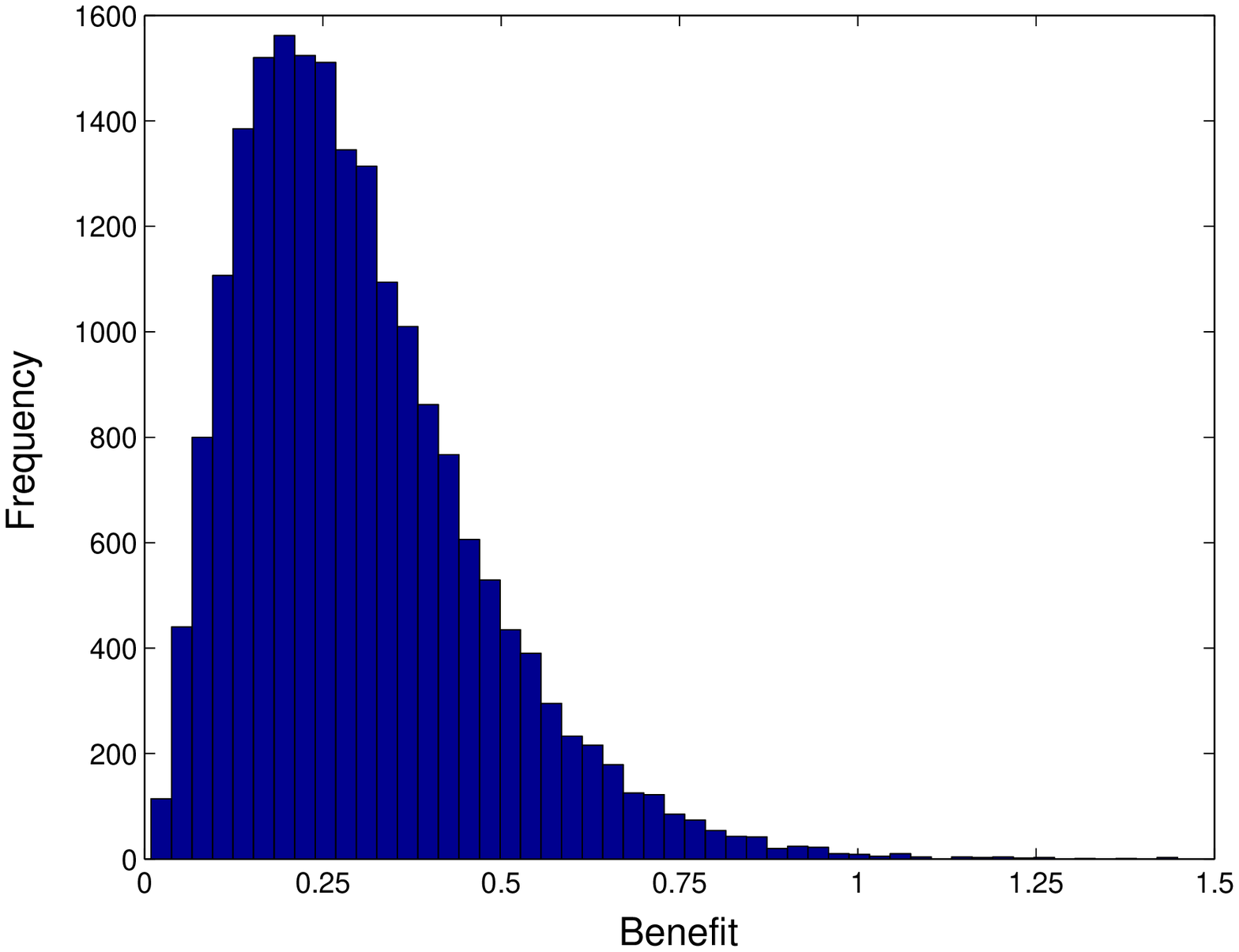}
    \includegraphics[width=0.35\textwidth]{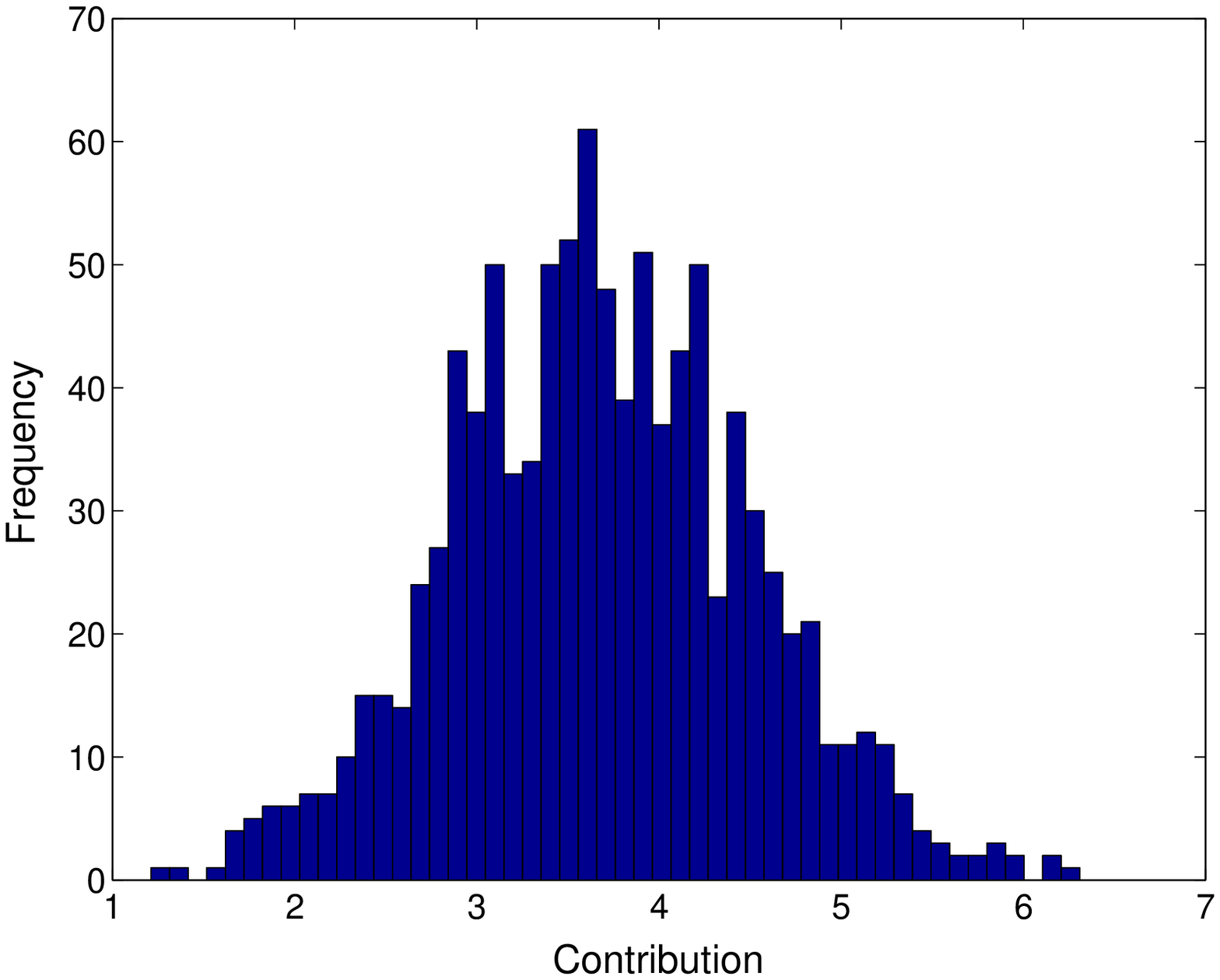}
  \end{center}
  \caption{Distributions of benefit and resulting contribution for
    1000 peers.  The
    histogram for benefits is the distribution of $b_{ij}$ for $b_{ij}
    \neq 0$.  Corresponding $b_\mathrm{av} =
    \frac{1}{N}\sum_{ij}b_{ij}$ is 6.0.  For contribution, the average
    is 3.68.}
    \label{hist-fig}
\end{figure}

In Figure \ref{hist-fig} we show the distribution of $b_{ij}$ and
$d_i$ for $N=1000$ peers.  The values of $b_{ij}$ were chosen from a
Gamma distribution such that $b_\mathrm{av} = 6.0$.  The equilibrium
values $d_i^*$ distribute themselves in a bell shaped distribution with
mean $d^*_\mathrm{av} = 3.68$.  If the system was completely
homogeneous, than the distribution of $b_{ij}$ would consist of a
single peak at $b=b_\mathrm{av}/(N-1)$ and the corresponding value of
$d^*_\mathrm{hi}$ from equation \ref{sym-equilibrium} would be 3.73
which is less than 1.5\% away from the value of $d^*_\mathrm{av}$.
\begin{figure}
  \begin{center}
    \includegraphics[width=0.35\textwidth]{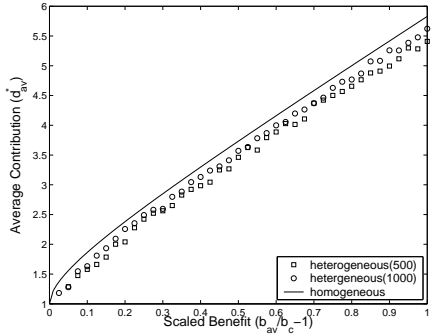}
    \caption{Average contribution at Nash equilibrium plotted against
    average benefit for 500 and 1000 peers.  The solid line is the
    prediction from the homogeneous system (equation
    \ref{eqbm-contrib-eqn}).}
    \label{equilibrium-fig}
  \end{center}
\end{figure}
In Figure \ref{equilibrium-fig} we show the equilibrium average
contribution by the peers as a function of average benefit.  The solid
line is the solution from the homogeneous system.  As expected, the
equilibrium contribution increases monotonically with increasing
benefit.  For average benefit $b_\mathrm{av} < b_c$, the iterative
algorithm converges to $d_i = 0$.  Note that the two sets of results
for 500 and 1000 peers almost coincide with each other.  So our
results are essentially independent of system size.

\begin{figure}
  \begin{center}
    \includegraphics[width=0.35\textwidth]{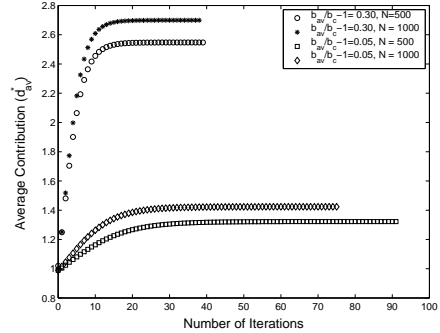}
  \end{center}
    \caption{Average contribution at Nash equilibrium plotted against
    number of steps required to reach Nash equilibrium for 500 and
    1000 peers.  We assumed convergence was reached when the average
    contribution per peer differed no more than one part in a million.
    The average initial value of contribution per peer is 1.0.  The
    values of average benefit are scaled as usual.}
    \label{convergence-fig}
\end{figure}

In Figure \ref{convergence-fig} we show the approach to convergence
for the learning algorithm.  The two data sets correspond to different
values of average $b_\mathrm{av}$.  Higher the average value of
$b_\mathrm{av}$, faster is the convergence to equilibrium.  As the
value of $b_\mathrm{av}$ approach the critical value $b_c$, approach
to equilibrium becomes slower and slower.  This is to be expected
since we have argued in section \ref{equilibrium-prop-sec} that near
the critical point, any deviation dies out very slowly.  We have
observed that for a wide set of initial conditions for $d_i$, the
process always converges to a unique Nash equilibrium.  For very small
initial values of $d_i$, we are close to the unstable Nash equilibrium
and the iteration converges to zero, i.e. the contribution by all
peers vanish and the system collapses.  The data for system collapse
is not shown, but Figure \ref{learning-fig} illustrates the situation.

\subsubsection{Inactive or Uncooperative Peers}
\begin{figure}
  \begin{center}
    \includegraphics[width=0.35\textwidth]{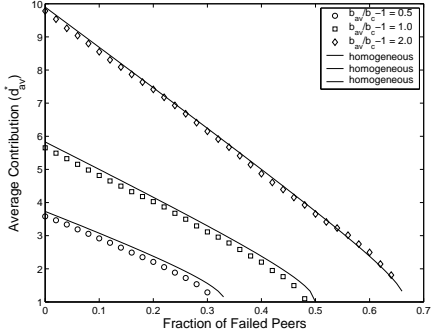}
    \caption{Average contribution at Nash equilibrium plotted against
    fraction of peers alive.  Total number of peers is 1000.  The
    solid lines are predictions from the homogeneous system model (equation
    \ref{eqbm-contrib-eqn}).}
    \label{kill-fig}
  \end{center}
\end{figure}

In Figure \ref{kill-fig}, we show the effect of some peers leaving
the system.  Intuitively one would think that if some peers leave the
system, the benefit per peer would be reduced and we should be seeing
pretty much the same behavior as in Figure \ref{equilibrium-fig}.  Our
simulations confirm this intuition.  As the fraction of active peers
dwindle, the contribution from each of the peers decrease and at some
point, the benefits are too low for the peers and the whole system
collapses.  The system can be pretty robust for high benefits : for a
benefit level of $(b_\mathrm{av}-b_c)/b_c = 2.0$, the system can
survive until 2/3 of the peers leave the system.  In contrast to
traditionally fragile distributed systems, we see that for P2P systems
robustness increase with size : as the system grows bigger and bigger,
benefits for each peer increases and the system becomes more robust to
random fluctuations.

\begin{figure}
  \begin{center}
    \includegraphics[width=0.35\textwidth]{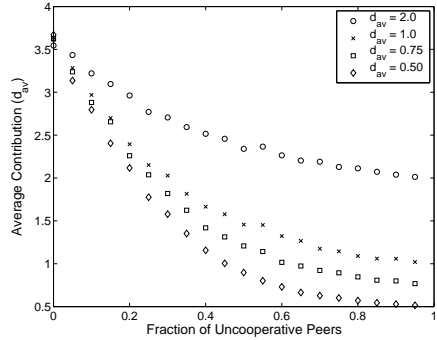}
    \caption{Average contribution at Nash equilibrium plotted against
    fraction of uncooperative peers.  Total number of peers is 1000.
    The labels specify the average contribution of uncooperative
    peers.   Average benefit is $b_{av}/b_c-1 = 0.5$.}
    \label{freeze-fig}
  \end{center}
\end{figure}

In Figure \ref{freeze-fig} we explore the effect of having peers which
behave uncooperatively, i.e. they refuse to adjust their contribution
and simply make a constant contribution.  The effect of such
non-cooperative peers is clear.  If they constitute 100\% of the
peers, of course the average contribution is equal to their
contribution.  Otherwise their effect is to bias the equilibrium
contribution value toward them.

\section{Discussion}
\label{impl-sec}
In this paper we have proposed a differential service based incentive
mechanism for P2P systems to eliminate free riding and increase
overall availability of the system.  We have shown that a system with
differential incentives will eventually operate at Nash equilibrium.
The strategy of a peer $P_i$ wishing to join the system depends on a
single parameter $b_i$ which is the benefit that $P_i$ can derive from
the system.  If the benefit $b_i$ is larger than a critical benefit
$b_c$, then the peer's best option is to join the system and operate
at the Nash equilibrium value of contribution.  If on the other hand
$b_i < b_c$, the peer is better off not joining the system.  When
$b_i=b_c$, the peer is indifferent between these two options.  These
properties are robust and do not depend on the details of the
particular incentive mechanism that is used.
\subsection{Implications for System Architecture}
The incentive policy that we have discussed can be implemented with
minor modifications to current P2P systems.  Let us look at some of
the modifications required.

Current P2P architectures do not restrict download in any way except
by enforcing queues and maximum number of possible open connections.
Our incentive scheme is easily implemented by accepting requests from
peers with a probability $p(d)$.  To prevent rapid fire requests from
the same peer, it will be necessary to keep record of a request for a
small duration of time.  In our discussion we have assumed that the
function $p(d)$ is same for all users, i.e. it is part of the system
architecture which can not be modified by users.  For greater
flexibility, it is possible to allow individual peers to configure
$p(d)$, but the effects on overall system performance is not clear.

The contribution is measured in terms of uptime and disk space.  When
a peer makes a request for a file, the contribution information can be
attached as an extra header to the request.  In fact, the current
Gnutella protocol already sends metadata like shared disk space and
uptime with its request messages.  New users can be given a default
value of contribution for a limited period of time so that they can
start using the system at a reasonable level.

There is incentive for peers to misreport contributions so that they
can reap the benefit of the system while making no contribution.  To
prevent such misuse, it is possible to implement a \emph{neighbor audit
scheme}.  Such a scheme is especially attractive in a fixed network
topology such as the CAN \cite{can} or Chord \cite{chord} system.
Every peer will continually monitor the uptime and disk space of its
neighbor.  If any doubt exists about the accuracy of the information
reported by a peer, the information can be verified from its neighbor.

\subsection{Alternative Metrics for Contribution and Incentive}
We have touched upon only a handful of questions that are relevant to
building a reliable P2P architecture with incentives.  There are many
unresolved issues which will have to be addressed in future by system
architects.  For example, what is the best metric for the contribution
of a user?  A popular metric is the number of uploads provided by a
peer.  So the peers that provide the most popular files and have
the highest bandwidth are deemed to contribute the most.  
Our metric, which simply
integrates disk-space over time does not discriminate against low
bandwidth peers or peers which provide file which are not very
popular.  Such a metric is very appropriate for a project like Freenet
\cite{freenet} which aspires to be an anonymous publishing system
regardless of the popularity of the documents published.  The metric that
is in practical use by KaZaA  is called
\textit{participation level} and is given by
\begin{equation}
\mathrm{participation\, level} = \frac{\mathrm{uploads\,in\,MB}}{\mathrm{downloads\, in\, MB}}\times 100.
\end{equation}
The participation level is capped at a maximum of 1000. 

Our analysis of incentives relied on the peers being rational and
trustworthy.  Trust is not easy to enforce.  The neighbor audit scheme
will deter individual misbehavior, but collusion among a set of peers
is still possible.  Another trust related problem involves malicious
peers who contribute fake files.  The idea of EigenTrust \cite{kamvar}
is a significant step in this direction which also protects against
collusion among malicious peers.

The incentive scheme we have outlined is through selective denial of
requests.  There are other ways to implement incentives.  For example
one could implement differential service for $P_i$ by restricting the
download bandwidth to a fraction $p(d_i)$ of the total bandwidth
available.  KaZaA's participation level operates on a similar
principle: if more than one peer requests the same file, the peer with
smaller participation level is pushed to the back of the queue.

Instead of implementing incentives on download level, one could also
restrict the search capabilities of a peer.  The basic idea is to
reduce the number of peers to which queries are propagated.  In
Gnutella, a peer forwards a query to its neighbors based on the Time
To Live (TTL) field.  By reducing the TTL of the query or by
forwarding the query only to a fraction of the total neighbors, the
search space for the query can be restricted.
\begin{equation}
  \mathrm{scaled\, TTL\, for\,} P_i = \lceil p(d_i) \times \mathrm{initial\, TTL} \rceil
\end{equation}
We note that the effect of restricting search using a function $p(d)$
is \textit{not} equivalent to restricting download using the same
function.  Network topology will have a significant role to play in
determining the actual set of files that a user has access to.
Regardless of the actual implementation of incentives, our conclusions
concerning existence and properties of the Nash equilibrium in the
system will remain qualitatively unchanged.

\bibliographystyle{plain}
\bibliography{p2p_model}
\end{document}